\documentclass[
    ,final            
  ]
  {aipproc}

\layoutstyle{6x9}
\usepackage{graphicx}
\usepackage{bm}

\def\A{{\bf A}}
\def\B{{\bf B}}

\begin{document}

\title{The Flux Tube Model: Applications, Tests, and Extensions}

\author{Eric S. Swanson}{
  address={Department of Physics and Astronomy, University of Pittsburgh, Pittsburgh PA, 15260}
}

\begin{abstract} 
A review and critique of the Isgur-Paton flux tube model of hadronic physics is presented.
This entails a detailed comparison with recent lattice gauge theory results which 
exposes both the successes and shortcomings of the model. Applications to hybrid masses, meson and
hybrid meson decays, hadronic charge radii, the spin-orbit force, baryonic hybrids, and hybrid 
photocouplings are also discussed. Finally, I comment on the issue of adiabatic surface crossing
which appears in both the flux tube model and lattice studies.

\end{abstract}

\maketitle

\section{Flux Tubes} 

The impending upgrade of CEBAF at Jefferson Lab and the advent of a new
experimental facility, Hall D,  devoted in large part to the exploration
of the gluonic excitations of QCD make it a propitious time to review the
flux tube model. The flux tube model (FTM) of Isgur and Paton is now nearly 20 years old\cite{IP}
and much has been learned in the meantime. This note shall review the 
foundations and predictions of the FTM and compare it, where possible, to lattice
gauge or other theoretical tests.  Extensions of the original ideas and new applications
are also treated.

The flux tube model is an attempt to form a tractable model of low energy QCD which
is based on longstanding ideas of string-like gluonic degrees of freedom. The central
idea is that gluons should rearrange themselves into flux  tubes
which, in the heavy quark limit, adjust their configuration instantaneously in response to
quark motion. Thus  quarks are constrained to move on adiabatic gluonic energy 
surfaces. The lowest such surface is the familiar `Coulomb+linear' potential of
the constituent quark model and lattice gauge theory. States based on this surface
thus form the `conventional' mesons of the quark model. Higher adiabatic surfaces
represent gluonic excitations and mesons built on these surfaces correspond to 
`hybrids'\cite{hyb}. Thus the flux tube model is a simple and intuitive way to extend the 
nonrelativistic constituent quark model to include gluonic degrees of freedom.

The model was originally motivated through a series of truncations on the
Euclidean time strong coupling QCD Hamiltonian with a lattice regulator (Hamiltonian
lattice gauge theory). The degrees of freedom are quarks fields on lattice sites and
gluonic `link variables' $U_\ell = \exp(-iag A_\mu(x))$ where $\ell$ represents the link ($x,\hat \mu$). 
In the strong coupling limit the Hamiltonian is given by

\begin{equation}
H_{scQCD} = {g^2\over 2a} \sum_\ell E_\ell^aE_{a\ell} + m\sum_n\bar \psi_n \psi_n
\end{equation}
where $g$ is the strong coupling, $a$ is the lattice spacing, 
and  $n$ is a lattice site. The velocity variables $\dot U_\ell$ have been replaced by
electric field operators $E_\ell$. 
Gauge invariant pure glue states are formed by closed (possibly
multiply connected) loops of link operators.
The commutation relation $[E^a,U] = T^a U$ then implies that the energy of these states is 
simply the sum of the quadratic colour charges of each link:

\begin{equation}
E_{\rm loop} = {g^2\over 2a}\sum_{\ell \in {\rm loop}} {\cal C}_\ell^2
\end{equation}
where ${\cal C}^2 = 4/3$ for a field in the 3 or $\bar 3$ representations, 10/3 for 6 or $\bar 6$, etc.
The presence of quarks permits gauge invariant states with open flux strings which terminate 
on quark colour sources or sinks. 
Perturbations to these states are provided by subleading quark hopping and magnetic
terms. The former allow flux tube breaking via quark pair production or quark motion.
The latter can change link colour representations, cause link hopping, or change 
loop topology.

Isgur and Paton simplify the dynamics by (i) assuming an adiabatic separation of
quark and gluon degrees of freedom (ii) neglecting `topological mixing' such
as loop breaking or loop Euler number changing transitions (iii) working in the nonrelativistic
limit.  The model is meant
to be applied to the `intermediate regime' where  $1/a \sim \sqrt{b}$.
They then model
link variable dynamics in terms of spinless colourless particles (`beads') of mass $ba$ where $b$ is
the string tension in the static quark potential. Finally these particles are assumed
to interact via a linear potential and perform small oscillations about their
resting positions. The result is a simple discrete string model for glue described
by the Hamiltonian:

\begin{equation}
H = b_0 R + \sum_n\left[ {p_n^2\over 2b_0a} + {b_0\over 2a}(y_n-y_{n+1})^2\right],
\end{equation}
where $y_n$ is the transverse displacement of the $n$th of $N$  string masses , $p_n$ is its
momentum, $b_0$ is a bare string tension, and $R=(N+1)a$ is the separation between the
static quarks. This Hamiltonian may be diagonalised in
the usual way yielding

\begin{equation}
H_{FTM} = b_0R + \left( {4\over \pi a^2}R -{1\over a} - {\pi \over 12R} + \ldots\right) + \sum_{n\lambda} \omega_n \alpha_{n\lambda}^\dagger \alpha_{n\lambda}
\end{equation}
where $\alpha_{n\lambda }^\dagger$ creates a phonon in the $n$th mode with polarization $\lambda$. Notice that the string tension has been renormalized by the first term in brackets. 
The last term in brackets is the L\"uscher term of string phenomenology\cite{L}. The
mode energies are given by $\omega_n = 2/a \sin[\pi n / 2(N+1)]$. Thus 
$\omega_1 \to \pi/R$ as $N\to \infty$ is the splitting between the ground state
Coulomb+linear potential and the first gluonic excitation surface at long range.
The energy of a given phonon state is approximately

\begin{equation}
E = E_0 + N {\pi\over R}
\label{E}
\end{equation}
with
\begin{equation}
N = \sum_{m=1}^\infty m( n_{m+} + n_{m-})
\end{equation}
where
$n_{m\pm}$ is the number of left (right) handed phonons in the $m$th mode.

\section{Hybrids} 

Hybrid mesons are constructed by specifying the gluonic states via phonon
operators and combining these with quark operators with a Wigner rotation matrix:

\begin{equation}
|LM_L; s \bar s; \Lambda, \{n_{m+},n_{m-}\}\rangle \propto \int d^3 r \varphi(r) D^L_{M_L\Lambda}(\hat r) \, b^\dagger_{r/2,s} d^\dagger_{-r/2,\bar s} \prod_m (\alpha_{m+}^\dagger)^{n_{m+}} (\alpha_{m-}^\dagger)^{n_{m-}} | 0 \rangle.
\end{equation}
The projection of the total angular momentum on the $q\bar q$ axis is denoted by $\Lambda = \sum_m (n_{m+}-n_{m-})$.
The parity and charge parity of these states are given by

\begin{equation}
P|LM_L; S M_S; \Lambda, \{n_{m+},n_{m-}\}\rangle = (-)^{L+\Lambda+1}|LM_L; S M_S; -\Lambda, \{n_{m-},n_{m+}\}\rangle,
\end{equation}

\begin{equation}
C|LM_L; S M_S; \Lambda, \{n_{m+},n_{m-}\}\rangle = (-)^{L+S+\Lambda+N}|LM_L; S M_S; -\Lambda, \{n_{m-},n_{m+}\}\rangle.
\end{equation}
States of good parity are thus formed as 
\begin{equation}
|LM_L; S M_S; \zeta; \Lambda, \{n_{m+},n_{m-}\}\rangle = {1\over \sqrt{2}}\Big(|LM_L; S M_S; \Lambda, \{n_{m+},n_{m-}\}\rangle + \zeta |LM_L; S M_S; -\Lambda, \{n_{m-},n_{m+}\}\rangle\Big)
\end{equation}
Possible single phonon $(m=1)$ mesons are listed in Table I.
Underlined quantum numbers represent {\it quantum number exotic hybrids} (mesons with 
quantum numbers not available to a $q\bar q$ state).

\begin{table}[h]
\begin{tabular}{cccc}
$\zeta$ &  $L$  &  $S$   &  $J^{PC}$ \\
\hline
+  & 1  &  0  & $1^{++}$ \\
+  & 1  &  1  & $(\underline{2},1,\underline{0})^{+-}$ \\
+  & 2  & 0  &  $2^{--}$ \\
+  & 2  & 1  &  $(\underline{3},2,\underline{1})^{-+}$ \\
\hline
-  &  1 &  0  & $1^{--}$ \\
-  & 1  & 1  & $(2,\underline{1},0)^{-+}$ \\
-  & 2  &  0  & $2^{++}$ \\
-  & 2 &  1  &  $(3,\underline{2},1)^{+-}$ \\
\end{tabular}
\caption{Some Single Phonon Mesons.}
\end{table}

Isgur and Paton obtained hybrid meson masses by solving a model Hamiltonian of quark
motion on the single-phonon excited surface:

\begin{equation}
H_{IP} = -{1\over 2 \mu} {\partial^2\over \partial r^2} + {L(L+1) - \Lambda^2 \over 2 \mu r^2} - {4 \alpha_s \over 3 r} + {\pi \over r}(1 - {\rm e}^{-f \sqrt{b} r}).
\label{HIP}
\end{equation}
The interaction term incorporates several important additional assumptions. Namely
the $\pi/r$ phonon splitting is softened at short range. The parameter $f$ which
appears in the softening function was estimated to be roughly unity\cite{IP}.
Furthermore, it was assumed that the attractive Coulomb ($1/r$) potential  remains
valid for hybrid mesons. Note that this is at odds with the expected 

\begin{equation}
V_{oge} = + {\alpha_s \over 6 r}
\label{Voge}
\end{equation}
potential of perturbative one gluon exchange (the colour factor is that appropriate to
gluon exchange between quarks in a colour octet state). This is an important assumption
which will be examined in more detail below. 

Finally the quark angular momentum operator is now complicated by the presence of 
gluonic/string
degrees of freedom. One may write

\begin{equation}
L_{q\bar q} = L - L_{S_{||}}- L_{S_\perp}
\end{equation}
where $L$ ($L_S$) is the total (string) angular momentum. Note that
$L_{S_\perp}$ mixes adiabatic surfaces. Using $L_{S_{||}} = \Lambda \hat r$ and
neglecting surface mixing  yields the centrifugal term  of Eq. \ref{HIP}.
This additional assumption will also be examined in the next section.

The hybrid masses obtained by solving Eq. \ref{HIP} are labelled $E_{IP}$ in Table II. 
Isgur and Paton also estimated
the effects of adiabatic surface mixing and used these as their final mass estimates
(labelled $E_{IP}'$). The column labelled KW is explained in the next section.
Finally, Table I implies that hybrids with quantum numbers
$2^{\pm \mp}$,$1^{\pm\mp}$,$0^{\pm\mp}$,$1^{\pm\pm}$ are all degenerate at this order in the
FTM.
\begin{table}[h]
\begin{tabular}{cccc}
flavour & $E_{IP}$  & $E_{IP}'$  &  $E_{KW}$\cite{KW}  \\
\hline
I=1  &  1.67 &  1.9  &  1.85 \\
I=0  &  1.67  & 1.9  &  1.85  \\
$s\bar s$  & 1.91  &  2.1 & 2.07  \\
$c\bar c$  & 4.19 &  4.3  & 4.34 \\
$b \bar b$ & 10.79  & 10.8 & 10.85 \\
\end{tabular}
\caption{Hybrid Mass Predictions}
\end{table}

The next section will present a survey of possible tests of this `zeroth order'
Flux Tube Model.

\section{Testing the Flux Tube Model} 

\subsection{Small Oscillation and Adiabatic Approximations}

The small oscillation approximation may be tested by considering a model of
transverse beads  interacting via a linear potential. Numerically solving such
a Hamiltonian\cite{BCS} reveals that the small oscillation approximation is accurate
for long strings but overestimates gluonic energies by an increasing amount as
the interquark distance shrinks. Typical energy differences are order 100 MeV at
1 fm. Similarly, the adiabatic approximation can be tested by numerically solving
the coupled quark-bead system. One finds\cite{BCS} that the adiabatic approximation
underestimates true energies by roughly 100 MeV, with slow improvement as the quarks
get very massive. It thus appears that these approximation errors tend to cancel 
each other, leaving the IP predictions intact.

\subsection{Gluonic Surfaces}

\begin{figure}[h]
\includegraphics[angle=-90,width=10cm]{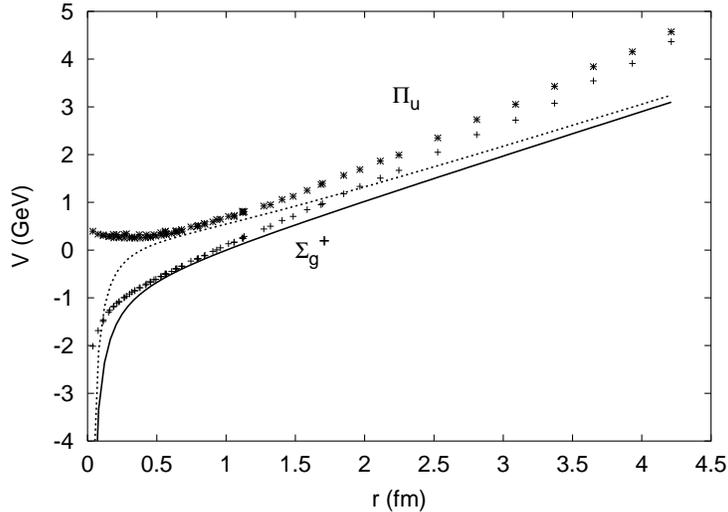}
\caption{Comparison of the Potentials of Eq. \ref{HIP} with lattice gauge theory\protect{\cite{JKM}}.}
\label{Vplot}
\end{figure}

Recent advances in computational speed and algorithms make it possible to test
some of the assumptions of the FTM against the predictions of lattice gauge theory.
Figure \ref{Vplot} shows the assumed ground state Coulomb+linear (solid line) and first
excited state (dashed line) potentials of Isgur and Paton. These are compared to lattice
Wilson loop 
computations of the same interactions (points)\cite{JKM}. It is apparent
that the IP potential overestimates the strength of the Coulomb potential and 
underestimates the string tension. Of course both of these model parameters  are
obtained by fitting the meson spectrum and therefore include 
effects which are not in the Wilson loop.

\subsubsection{Should the Coulomb Potential Appear?}

More troubling is the first
excited potential (bursts) which shows signs of saturating (or perhaps turning repulsive) at
small distances, in disagreement with the assumption of Isgur and Paton. Indeed,
simply omitting the attractive Coulomb portion of the IP hybrid potential 
dramatically improves agreement of the model with the lattice.  

\begin{figure}[h]
\includegraphics[angle=-90,width=10cm]{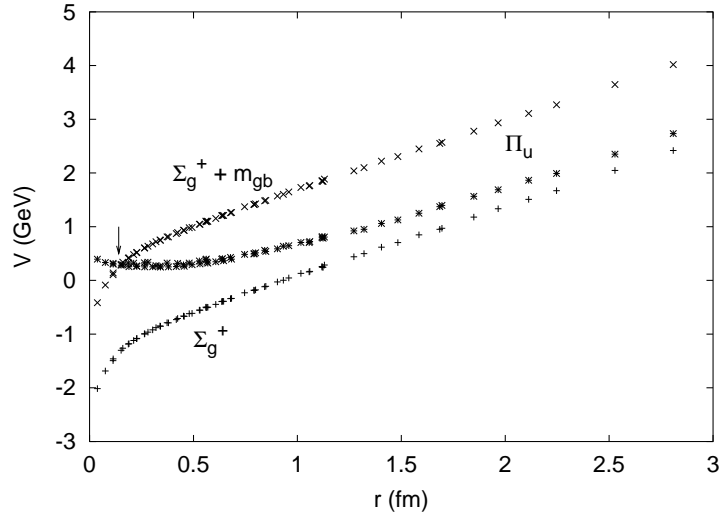}
\caption{Adiabatic Surface Crossing.}
\label{vcross}
\end{figure}

This disagreement represents something of a conundrum because Isgur and Paton had a good
physical reason to employ the colour singlet $q\bar q$ Coulomb interaction in their
model: if one assumes the repulsive short range interaction of Eq. \ref{Voge},
it becomes energetically favourable for the system to emit a gluon 
once the interquark separation becomes small enough.
This gluon combines with the `valence' gluon of the hybrid
to form a scalar glueball, thereby changing the quark colour configuration to that
of a singlet, and the Coulomb potential to the attractive form of Eq. \ref{HIP}. The
point at which this should happen is indicated by the arrow in Fig. \ref{vcross}.
It is clear that the lattice sees no such behaviour.  This may be simply because
the minimal relative momentum permitted on the lattices employed in the study 
were too coarse to permit the decay. 

A physical reason for the suppression of this
coupling is also possible. For example,  if one considers the hybrid to be dominated
by a Fock space component consisting of a constituent quark, antiquark, and gluon, then
the postulated transition occurs through gluon emission from either the valence gluon
or a valence quark. In the former case the coupling to a glueball is 
zero due to the colour overlap while in the latter case the coupling is suppressed by
the large (infinite) quark mass. Thus one expects the coupling between the surfaces to
be very small, which implies that the surface mixing will not be seen unless lattices
with exceptionally large temporal extents are employed. 

Which short distance behaviour should the model use? It seems clear that surface
mixing which changes Fock sectors should not be used in a  potential model. Rather,
such mixing should be incorporated by explicitly including the appropriate
transition operator in the formalism. Thus, for example, flux tube breaking
is expected to eliminate the linear potential for distances beyond roughly one
fermi when light quarks are present. However, such a truncated linear potential
should not be used to construct mesons, rather one should employ the linear potential
to form a basis of bound states and then include mixing to four quark states in
an appropriate form. This approach avoids the unpleasant  situation of having a
mesonic ionization energy. Similarly, the hybrid potential should not include the
mixing term since this mixes hybrid with hybrid+glueball Fock states.

\subsubsection{Higher Surfaces}

\begin{figure}[h]
\includegraphics[width=10cm]{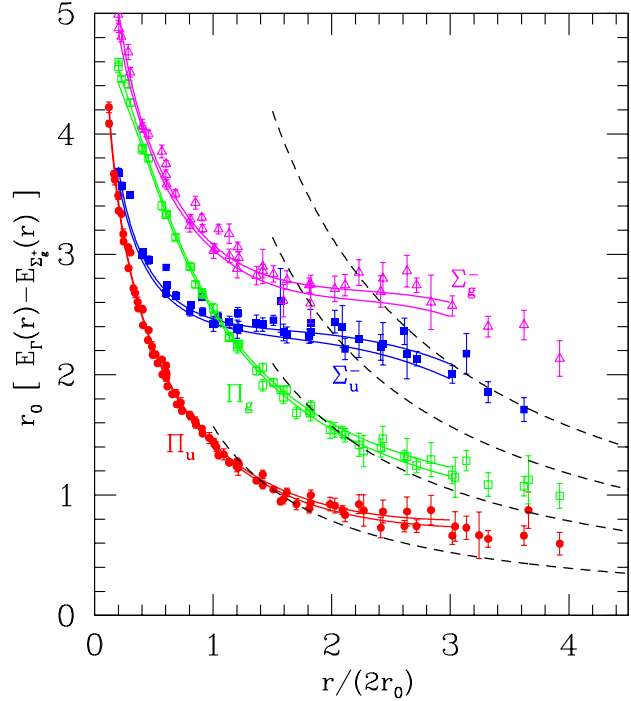}
\caption{ Hybrid Surface Energy Differences. 2$r_0$ is roughly 1 fermi\protect\cite{JKMstring}. }
\label{strings}
\end{figure}

The FTM predicts an entire tower of hybrid surfaces and it is interesting to
examine their form in light of lattice data.
Juge, Kuti, and Morningstar have carried out a detailed analysis of the relationship of the hybrid surfaces of  Fig. 1  to the string excitations of Eq. \ref{E}\cite{JKMstring}. They have found
that surface excitations only have $\pi/r$ splittings for very large source separation
(roughly 4 fermi or greater). This is shown in Fig. \ref{strings} where the dashed lines are
the predicted $N \pi/r$ energy differences of the FTM. This is something of a surprise since one
expects a phonon-like excitation spectrum on general grounds. It appears that QCD strings are 
complex objects at intermediate distance scales.
Finally, the figure shows a cross over region at about 1 fermi
where the surfaces move from a perturbative behaviour (characterized by the `gluelump' spectrum) to a
more string-like behaviour. 

\subsubsection{Hybrid Masses Revisited}

In light of these issues it is appropriate to revisit the original hybrid mass
estimates of Table I. A number of variants of the IP calculation were made in Ref. \cite{KW}.  Before presenting some of these, note that there is additional ambiguity in the
quark angular momentum term of Eq. \ref{HIP}, namely it is also possible to set
$L_{q\bar q} = L - J_g$
where $J_g$ is the total gluonic angular momentum. Squaring then yields 

\begin{equation}
L_{q\bar q}^2 = L(L+1) - 2\Lambda^2 + \langle J_g^2\rangle.
\label{cent2}
\end{equation}
It has been found that setting $J_g$ = 2 yields good agreement with lattice data\cite{JKMJg}(this was also found in a constituent quark model\cite{SS2}). Notice that this is
{\it not} numerically the same as the centrifugal term in Eq. \ref{HIP}.

Employing the hybrid lattice potential of Fig. 1 and Eq. \ref{cent2} yields the hybrid
masses labelled $E_{KW}$ in Table II. It is seen that these can differ by up to 
200 MeV from the IP predictions of the second column. The similarity to the third 
column is a fluke since adiabatic surface mixing is not included in this computation.
Merlin and Paton have noted that the majority of adiabatic mixing effects may be 
absorbed into
the static potential by including the moment of inertia of the string in the centrifugal
term (see the discussion below):

\begin{equation}
{1 \over 2 \mu r^2} \to {1\over 2\mu r^2 + {1\over 6} b r^3},
\label{cent3}
\end{equation}
and with  a more important effect which modifies the strength of the $\pi/r$ splitting
to be larger as $r$ becomes larger than $m_q/b$. Thus the final revisited  prediction for
a light hybrid is roughly 2.1 GeV.

\section{The IKP Decay Model} 

Shortly after its introduction, the
flux tube model of meson structure was extended by Isgur, Kokoski, and Paton to
provide a description of meson\cite{IK} and hybrid\cite{IKP} decays. The 
transition operator was envisioned as arising due to the quark hopping term of the
lattice QCD Hamiltonian. The lowest terms in the expansion of this operator are 

\begin{equation}
H_{hop} = \sum_{n,\mu} \psi_n^\dagger \alpha\cdot \mu \psi_n  + a \sum_{n,\mu}\psi_n^\dagger\alpha\cdot\mu\nabla\cdot\mu \psi_n.
\end{equation}
If one assumes a smooth string then the first term dominates as the lattice spacing gets
small and one has a $^3S_1$ strong decay operator. Alternatively, if the string is rough
then the first term averages to zero upon summing over all local string orientations
and the second term dominates, yielding a  $^3P_0$ strong decay operator. The authors
of Ref. \cite{IK,IKP} assume the second scenario since it is supported by experiment\cite{GS}.

Flux tube degrees of freedom were incorporated by assuming factorization:

\begin{equation}
\langle \{\ldots\} b d; \{\ldots\} b d | {\cal O} | \{\ldots\} b^\dagger d^\dagger\rangle \approx \langle bd; bd| ^3P_0 | b^\dagger d^\dagger\rangle \cdot
\langle \{\ldots\}; \{\ldots\} | \{\ldots\}\rangle
\end{equation}
The first matrix element on the right hand side is a typical $^3P_0$ mesonic
decay overlap. The second represents the overlap of the gluonic/flux tube degrees
of freedom. Assuming that the quark pair creation occurs at a transverse distance
$y_\perp$ from the interquark axis of the parent meson yields the results

\begin{equation}
\langle \{0 \ldots \};\{0 \ldots \}| \{ 0 \ldots \}\rangle \sim {\rm e}^{-f b y_\perp^2}
\end{equation}
for meson decay and
\begin{equation}
\langle \{0 \ldots \};\{0 \ldots \}| \{ 1 \ldots \}\rangle \sim y_\perp{\rm e}^{-f b y_\perp^2}
\end{equation}
for hybrid decay. The factor  $f$ is a computable constant of order unity.  The extra 
factor of $y_\perp$ in the hybrid decay vertex forces the decay to pairs of identical
$S$-wave mesons to be zero. This is the origin of the famous `S+D' selection rule in 
this model (it occurs in other models for different reasons).

Unfortunately, it will be some time before lattice gauge theory has progressed to the
point where models such as these can be thoroughly tested. However, preliminary 
computations reveal that a substantial closed flavour decay mode (such as 
$b\bar b g \to \chi_b \eta$) may exist\cite{UKQCD}. In the meantime we hope that
experiment will provide some clues. An alternative decay model is discussed in 
the next section.

\section{Extensions} 

The hadronic decay model of the proceeding section is its most well known
extension. However, the model has been extended in several other directions
as well; some of these are described here.

\subsection{Charge Radii and Hybrid Decays} 

Several years ago Isgur pointed out that the energy carried by the  flux
tube will change several features of the naive quark  model\cite{I} (see also \cite{MP2}). 
For example,
zero point
oscillation of the flux tube about the interquark axis will induce transverse
fluctuations in the quark positions, something which is not present when the
flux tube is treated as potential. The additional fluctuations have the effect of increasing the
 charge radius of a heavy-light meson (qQ):

\begin{equation}
r_Q^2 = \left[ \left( {m_q \over m_q+m_Q}\right)^2 + {2 b \over \pi^3m_q^2}\zeta(3)\right] \langle r^2\rangle
\end{equation}
where the second term in the bracket is the new contribution.
He estimated this to give rise to a 50\% increase in charge radii of light quark hadrons.

This observation has subsequently been expanded upon by Close and Dudek who
observe that radiative decays of hybrid mesons may proceed because the recoil of the 
radiating quark affects the string degrees of freedom giving a nonzero overlap of
the flux tube wavefunction  with the ground state flux tube wavefunctions of 
ordinary mesons. Preliminary computations with this mechanism have appeared\cite{CD}.

A similar scheme involving the emission of pointlike pions may be used to compute hybrid
decays to final states such as $\pi\rho$\cite{CD2}. The most striking result here is that
this decay mechanism evades the `S+D' suppression discussed above.

\subsection{Adiabatic Surface Mixing} 

Merlin and Paton examined the effects of adiabatic surface mixing on the leading
order FTM by considering the complete quark-bead system {\it ab initio}\cite{MP}.
Although the effects can be quite complicated, with mixing to all surfaces possible,
they found that the majority of the effects can be absorbed in a redefinition of the
hybrid potential by including the rigid body moment of inertia of the string in the
centrifugal term and by modifying the strength of the $\pi/r$ term (see Eq. \ref{cent3}). 

An explicit computation revealed
found mass shifts of order -100 MeV for conventional S-wave light quark mesons and +200
MeV for light quark hybrids. 
The resulting mass splittings are quoted in column three of Table I and were
used by Isgur and Paton to form their final estimates of the lowest lying hybrid masses.

\subsection{Spin Orbit Forces I} 

Merlin and Paton also examined spin orbit forces in the context of the FTM\cite{MP2}.
The idea was to map the operators of the leading spin orbit term in the heavy quark
expansion  of QCD, namely $V_{SO} = g/2m\, {\bm \sigma}\cdot \B$, onto FTM 
degrees of freedom (phonons).
Merlin and Paton did this by identifying the magnetic field with the lattice 
operator

\begin{equation}
T^a B^a(x) \sim {1\over 2 ga^2} \left( U_P(x) - U_P^\dagger(x)\right).
\end{equation}
Since the plaquette operator moves flux links in a fixed topological sector, it is
natural to identify the magnetic field with the bead kinetic energy. Doing so then allows
one to write $V_{SO}$ in terms of phonons.

Explicit computations revealed that spin orbit splittings due to $V_{SO}$ are 
small and that the majority of the splittings arise from Thomas precession,
$V_{Th} = {1\over 4} (\ddot {\bm r}_q \times \dot {\bm r}_q)\cdot {\bm \sigma}$. This
is modelled by including the effects of phonons on the quark coordinate (see the
discussion of the charge radii above).  The mass splittings for light hybrids are
listed in Table III.
One sees that the lowest member of the octet of light hybrids is predicted to be 
the $2^{+-}$ 
while the heaviest is the $0^{+-}$. This appears to be in conflict with lattice gauge
theory which finds that the lightest hybrids are $1^{-+}$.

\begin{table}[h]
\begin{tabular}{l|llllllll}
$J^{PC}$ & $2^{+-}$ & $2^{-+}$ & $1^{-+}$ & $0^{-+}$ & $1^{+-}$ & $0^{+-}$ & $1^{++}$ & $1^{--}$ \\
\hline
$\delta M$ (MeV) & -140 & -20 & 20 & 40 & 140 & 280 & 0 & 0 \\
\end{tabular}
\caption{Spin Orbit Hybrid Mass Splittings\cite{MP2}.}
\end{table}

\subsection{Spin Orbit Forces II} 

Spin-dependent forces in the FTM were also taken up by the authors of
Ref. \cite{SS4} in an attempt to resolve a conundrum in the spin-orbit sector of
the quark interaction. The issue is that many models of hadrons prefer a vector
Dirac structure of confinement, rather than the phenomenologically accepted Dirac
scalar interaction. This may be studied by using  the heavy quark Foldy-Wouthyusen
version of the QCD Hamiltonian in Coulomb gauge. The resulting ${\cal O}(1/m)$ and 
${\cal O}(1/m^2)$ operators, $H_1$ and $H_2$, depend on the chromoelectric and magnetic fields 
and hence are nonperturbative. In a similar approach to that of Merlin and Paton,
the authors of Ref. \cite{SS4} chose to study these operators by
mapping the chromofields to phonons. In this case the mapping was 
based on the idea that the electric field counts string length and therefore should be
mapped as $E_\lambda^a(x=na) = {\sqrt{b_0}\over a^2}\left(y^a_\lambda(n+1) - y^a_\lambda(n)\right)$
where $a$ is a colour index and $\lambda$ is a polarization index. Imposing the canonical 
commutation relations gives the magnetic field as a derivative with respect to transverse 
displacement. Finally both expressions can be mapped to phonon degrees of freedom with the
standard Fourier transform. Notice that this approach has the flux tube beads carrying colour
charge.

The form of the effective spin-dependent interactions were then evaluated by inserting these
field operators into $H_1$ and $H_2$. It was found that an effective scalar spin orbit
interaction could indeed arise in the flux tube picture of hadronic structure even though the
static confining interaction was vector. This was the case because nonperturbative mixing of
mesons with hybrids contribute to spin-dependent interactions and can change the naive
expectations based on the nonrelativistic reduction of a simple interaction kernel.

\subsection{Vector Decay Model} 

The same chromofield--phonon mapping which was used in the study of the spin orbit
interaction\cite{SS4} was used to examine hybrid decays in Refs. \cite{PSS}. In this case the operator
of interest is simply $V = -g\int \bar \psi {\bm \alpha}\cdot \A \psi$. The 
resulting decay vertex is given by

\begin{equation}
H_{int} = {i g a^2 \over \sqrt{\pi}}  \sum_{m,\lambda} \int_0^1 d \xi
\cos(\pi \xi)  T^a_{ij}\, b^\dagger_i(\xi {\bf r}_{Q\bar Q})
{\bm\sigma}\cdot \hat{\bf e}_{\lambda}(\hat{\bf r}_{Q\bar Q}) \left(\alpha^a_{m \lambda}
 - \alpha^{a \dagger}_{m \lambda} \right) d^\dagger_j(\xi {\bf r}_{Q\bar Q}),
\end{equation}

\noindent
where the $\hat{\bf e}(\hat{\bf r})$ are polarization vectors orthogonal to $\hat {\bf r}$.
The integral is defined along the $Q\bar Q$ axis only. This model also has an `S+D'
decay rule, however in contrast to the IKP decay model this is due to a node along the
interquark axis (the $\cos(\pi\xi)$) which causes  the selection rule, rather than a node
perpendicular to the interquark axis.

The phenomenology of this model has been presented in the second of Ref. \cite{PSS}. One
finds that it is similar to a $^3S_1$ model of meson decays in that $D$ wave decay modes tend
to be suppressed. Recent comparison with experiment have proven surprisingly accurate and
lend support to hybrid interpretations for nonexotic $2^{-+}(2003)$ and $1^{++}(2096)$
states\cite{JK}.

\subsection{Hybrid Baryons}

Capstick and Page have made a detailed study of baryon flux tube dynamics\cite{CP}.
This is a technically challenging problem due to the multitude of vibrational and rotational
modes which are available to a Y-shaped string system. However, they have found that the
problem simplifies considerably because the string junction decouples to good accuracy 
from the rest of the bead motion. Thus a hybrid baryon can be approximated by three 
quarks coupled via linear potentials to a massive `junction bead'. The dynamics of this
system are completely specified by the FTM and variational calculations indicate that the
lowest  lying hybrids are $J^P = {1\over 2}^+$ and ${3\over 2}^+$ states at approximately
1870 MeV.

Happily, lattice investigations of the static baryon interaction have begun\cite{BFT}. The chief
point of interest is whether the expected flux tubes form into a `Y' shape or a 
`$\Delta$' shape.
This may be addressed by carefully examining the baryonic energy in
a variety of quark configurations.  
Current results are mixed, with some groups claiming support for the two-body hypothesis
\cite{Alexandrou} and some for the three-body hypothesis\cite{Suganuma}.  Finally, a strong
operator dependence in the flux tube profiles has been observed\cite{OW}, which clearly 
needs to be settled before definitive conclusions can be reached.

\section{Conclusions} 

The flux tube model is now nearly 20 years old. In this time it has been applied to
an increasing array of problems and extended in several directions, chiefly by taking
seriously the idea that dynamical string-like gluonic degrees of freedom are important in
the low lying mesons. A number of additional extensions are: 

(i) glueballs (glue loops). A
preliminary study has been made in the original paper\cite{IP}, however much remains to be
done here.  Comparison with lattice gauge theory will provide a crucial test.

(ii) FT effects in baryons. The charge radii ideas of Isgur, Close, and Dudek have immediate
impact on baryons and should be studied.

(iii)  adiabatic surfaces. It is worthwhile to attempt to leverage new precise lattice data on
the gluelump spectrum and the hybrid adiabatic surfaces to improve the FTM in detail.

(iv) the FTM may allow one to improve the semiclassical fragmentation formalism\cite{IP}.

(v) long range spin-spin and spin-orbit forces should be re-examined in an attempt to 
pin down this difficult aspect of nonperturbative QCD.

In summary, the FTM provides a compelling picture of strong QCD dynamics; however, it is a picture only!
We have seen that the FTM correctly describes the level orderings and, perhaps, splittings of
gluonic adiabatic energy surfaces at large distances (perhaps as large as 4 fm). The model
fails to describe the spectrum at small interquark distances (although, of course,
it can be amended). Furthermore, the original IP model of hybrids is likely to 
be incorrect in 
many details, although its phenomenology may be surprisingly robust. We have seen that it
is possible to extend the model in many different ways. Of course these extensions 
rely on detailed aspects of the FTM which are untested at best. It appears likely, for 
example, that the spin orbit splitting of Ref. \cite{MP2} do not agree with lattice data.
In the end, the utility of a tractable and appealing model should not be underestimated.

\begin{theacknowledgments}

I am grateful to the organizers of the JLab/INT  Workshop on Gluonic Excitations  for
the invitation to review a topic which I have been following for nearly
two decades.
This work was supported by the DOE under contracts DE-FG02-00ER41135  
and DE-AC05-84ER40150.

\end{theacknowledgments}



\bibliography{sample}

\end{document}